\documentclass[prd,nofootinbib,twocolumn,superscriptaddress]{revtex4-1}
\usepackage{amssymb}
\usepackage{amsmath}
\usepackage{graphicx}
\usepackage{color}

\usepackage[pdftitle={Comparisons of eccentric binary black hole
    simulations with post-Newtonian
    models},pdfauthor={Ian Hinder, Frank Herrmann, Pablo Laguna,
    Deirdre Shoemaker}]{hyperref}

\begin{document}

\title{Comparisons of eccentric binary black hole simulations 
with post-Newtonian models}

\author{Ian Hinder}
\affiliation{Max-Planck-Institut f\"ur Gravitationsphysik,
  Albert-Einstein-Institut, Am M\"uhlenberg 1, D-14476 Golm, Germany}

\author{Frank Herrmann}
\affiliation{Department of Physics, University of Maryland, College
  Park, MD 20742, USA}

\author{Pablo Laguna}
\affiliation{Center for Relativistic Astrophysics, School of Physics,
  Georgia Institute of Technology, Atlanta, GA 30332, USA}

\author{Deirdre Shoemaker}
\affiliation{Center for Relativistic Astrophysics, School of Physics,
  Georgia Institute of Technology, Atlanta, GA 30332, USA}

\begin{abstract}
  We present the first comparison between numerical relativity (NR)
  simulations of an eccentric binary black hole system with
  corresponding post-Newtonian (PN) results.  We evolve an equal-mass,
  non-spinning configuration with an initial eccentricity $e \approx
  0.1$ for 21 gravitational wave cycles before merger, and find
  agreement in the gravitational wave phase with an adiabatic
  eccentric PN model with 2~PN radiation reaction within $0.1$ radians
  for 10 cycles.  The NR and PN phase difference grows to 0.7 radians
  by 5 cycles before merger.  We find that these results can be
  obtained by expanding the eccentric PN expressions in terms of the
  frequency-related variable $x = (\omega\,M)^{2/3}$ with $M$ the
  total mass of the binary. When using instead the mean motion $n = 2
  \pi /P$, where $P$ is the orbital period, the comparison leads to
  significant disagreements with NR.  
\end{abstract}

\maketitle

\section{Introduction}

Tremendous progress towards detecting gravitational waves is being
made by observational efforts such as LIGO, VIRGO and GEO600.  Just
recently, LIGO has reached its designed sensitivity and is currently
undergoing enhancements to increase the sensitivity by an order of
magnitude as a step towards advanced LIGO. In anticipation of these
enhancements, it is essential to have models of gravitational
waveforms for all sources of gravitational radiation, in particular
for binary black hole systems, since they are expected to be one of
the most promising
sources~\cite{Barish:1999vh,Waldmann:2006bm,Hild:2006bk,Acernese:2002bw,Acernese:2006bj}.

Constructing these waveforms is a non-trivial task. Generating a
complete waveform involves numerically solving the full Einstein
equations in order to correctly describe the last few orbits and
merger.  This is computationally intensive with simulations running
for weeks to produce a single accurate waveform.  Furthermore, the
parameter space of merging black hole binaries is quite large. In
addition to the intrinsic black hole parameters (masses, spin
magnitudes and orientations), there are the orbital parameters
(eccentricity and semimajor axis). Because of the computational cost
of producing numerical waveforms, the only way to have a hope of
covering the parameter space efficiently is to use
waveforms that combine NR solutions with results from the PN
approximation.  To achieve this goal, it is firstly important to
cross-check the two methods to ensure that they give compatible
results where PN is valid, namely for large enough binary separations.
Secondly, it is necessary to investigate how close to the merger one
can use the PN results (a recent study~\cite{Yunes:2008tw} addresses 
this question in the extreme-mass-ratio case using the theory of
optimal asymptotic expansions).
 
Due to recent
advances~\cite{2005PhRvL..95l1101P,Campanelli:2005dd,Baker:2005vv} in
the field of numerical relativity, long-term accurate and stable
evolutions of binary black hole systems spanning several orbits as
well as the merger are now possible.  The initial separations of the
black holes in these simulations are now sufficient to make
comparisons with waveforms generated in the PN
approximation. Such comparisons have been made for equal-mass
non-spinning~\cite{Baker:2006yw,2007PhRvD..75l4018B,Baker:2006ha,Hannam:2007ik,
  Gopakumar:2007vh,Boyle:2007ft,Damour:2007vq,Boyle:2008ge},
unequal-mass~\cite{Berti:2007fi}, and spinning~\cite{Hannam:2007wf}
binaries, all in \emph{quasi-circular} orbits. As a result, hybrid
waveforms for quasi-circular orbits have been constructed which
combine the PN waveform, accurate when the black holes are far apart,
with the late inspiral and merger waveform that can only be obtained
using full
NR~\cite{Ajith:2007kx,Ajith:2007xh,Buonanno:2007pf,Pan:2007nw}.

In this paper, we take the next step and extend for the first time the
NR and PN comparison to the case of \emph{eccentric} binary black hole
systems.  Only recently the first NR studies of bound eccentric binary
black hole systems have been performed \cite{Sperhake:2007gu,Hinder:2007qu}, where
the dependence of the final
black hole mass and spin on the initial eccentricity for non-spinning,
equal-mass systems was studied.
It has long been known
that far-separated eccentric binary systems emitting gravitational
radiation will circularize~\cite{Peters:1964zz}.  However, it was not
known what would happen if a binary black hole system still had
significant eccentricity in the late stages of inspiral.  Rather than
forming a final black hole with a higher or lower spin, the NR results
showed that even up to an initial eccentricity of $e \sim 0.4$, the
final black hole mass and spin were the same as in the circular case,
indicating that the rate of loss of eccentricity was sufficient for
the binary to circularize prior to or during merger.  Due to the
tendency of eccentric binary systems to circularize, most of the
expected astrophysical binary black hole sources for Earth-based
gravitational wave detectors will have lost all their eccentricity by
the time their waves enter the frequency band of the
detector. However, several astrophysical scenarios have been proposed
in which binary black hole systems in eccentric orbits might be
detectable, for which it will be necessary to understand the dynamics
and waveforms of eccentric binary black hole systems.

One such scenario may occur in the dense cores of globular clusters,
where interactions between pairs of binary black hole systems eject
one of the black holes, resulting in a stable {\em hierarchical
triple}.  This is a three-body system (three-body black hole systems
have also been studied recently in NR~\cite{Campanelli:2007ea})
consisting of two closely bound black holes and a third orbiting the
center of mass of the first two.  When the two orbital planes are
strongly tilted with respect to each other, tidal forces from the
third body can cause an orbital resonance, increasing the eccentricity
of the inner binary.  This is known as the Kozai
mechanism~\cite{1962AJ.....67..591K}.  It has been
suggested~\cite{2003ApJ...598..419W} that this could lead to
eccentricities greater than about 0.1 at the time the binary enters
the frequency band of advanced ground-based detectors, followed by
merger driven by gravitational radiation reaction.  Stellar mass black
hole binaries in globular clusters are expected to have a thermal
distribution of eccentricities~\cite{Benacquista:2002kf}, and
intermediate mass black holes in globular clusters are expected to
have eccentricities between 0.1 and 0.2 while they are in the
frequency band of the LISA detector~\cite{Gultekin:2004pm}.
Supermassive black holes are also sources for LISA, and it is
currently unknown what eccentricity they might
have~\cite{1976ApJ...204L...1T}. They could potentially merge within
the Hubble time from highly eccentric orbits if the Kozai effect was
occurring~\cite{Blaes:2002cs}.  It has also been shown that massive
black hole binaries in disks of gas can merge without losing
eccentricity if the disc is rotating in the opposite sense to the
binary orbit~\cite{Dotti:2006dc}.  Being able to measure the non-zero
eccentricity from the waveforms will tell us about the physics of the
system, and may also have implications for detection if quasi-circular
templates are used.

The energy and angular momentum fluxes from the gravitational waves
emitted by a comparable mass eccentric binary were originally
determined by Peters and Mathews~\cite{PhysRev.131.435,Peters:1964zz} in
the Newtonian limit.  By balancing the time-averaged far-zone fluxes
of energy and angular momentum with the loss of binding energy and
angular momentum in the orbit, the rate of decay of the orbital
semimajor-axis and eccentricity could be determined in the {\em
adiabatic} approximation.  The result showed that the eccentricity of
a binary reduces by approximately a factor of three when the semimajor
axis is halved.

The next order corrections to this result were obtained 
to 1~PN and 1.5~PN order, enabling the study of the evolution of the
orbital elements using the {\em quasi-Keplerian} parametrization of
the orbit~\cite{Wagoner76,Blanchet:1989cu,1992MNRAS.254..146J,Blanchet:1993ec,Rieth:1997mk}. 
With the use of a generalized Keplerian
representation~\cite{Damour:1988mr,Schaefer93,1995CQGra..12..983W}, this
work was extended to 2~PN~\cite{Gopakumar:1997bs,Gopakumar:2001dy}.

An improved {\em method of variation of
constants} has been developed~\cite{Damour:2004bz,2006PhRvD..73l4012K}
in order to construct models which for the first time go beyond the
adiabatic approximation.  Very small oscillations in the
orbital elements were found on the timescale of the orbital period.  The
conservative 3~PN dynamics of an eccentric system in the
quasi-Keplerian representation have been derived~\cite{Memmesheimer:2004cv}.
Recently, the complete 3~PN energy
and angular momentum fluxes have been
determined~\cite{Arun:2007sg, Arun:2007rg, ArunPhD, arunblanchetam}.

The availability of the energy flux to 3.5~PN
order~\cite{2006LRR.....9....4B} in the quasi-circular case has led to
successful matches with NR waveforms, with agreement in the waveform
phase within 0.05 radians between 30 and 15 cycles before merger, and
within several radians up to the merger~\cite{Boyle:2007ft} for some
PN models, though the level of agreement near merger is model dependent.
The TaylorT4 model, specifically,
agrees within 0.05 radians up to $M \omega_\mathrm{gw} = 0.1$.
Recently, it has been shown that for TaylorT4,
the energy flux is identifiably different from the NR
result even 25 cycles before merger \cite{Boyle:2008ge}.

A circular binary black hole inspiral gives rise to waveform dynamics
which are in some sense simple: the amplitude and frequency increase
monotonically, which may explain why the adiabatic approximation works
so well.  Eccentric orbits on the other hand give rise to waveforms
with oscillations in the amplitude and frequency, and comparison with
NR in this case will provide a significantly more stringent test of
the PN approximation.

In this paper, we present the first analysis of the agreement between
PN and NR eccentric waveforms.  We restrict to the equal-mass,
non-spinning case.  We use the 3~PN conservative quasi-Keplerian orbit
equations~\cite{Memmesheimer:2004cv}, combined with the 2~PN evolution
of the orbital elements~\cite{Damour:2004bz} to construct adiabatic PN
waveforms, determined by four independent initial parameters.  We then
present a full NR evolution which starts 21 gravitational wave cycles
before merger with an estimated initial eccentricity of $e \approx
0.1$.  We assume that the NR simulation gives the final stage of a full
waveform, such as one that would be observed in nature.  We then
choose a fitting interval in time and use least squares fitting to
find the parameters of the PN waveform which best matches the
numerical data in that interval.  We find agreement
between the NR and PN gravitational wave phase within $0.1$ radians
for 10 wave cycles at the start of the simulation.  The NR and PN phase
difference grows to 0.6 radians 5 cycles before merger, corresponding
to $M \omega_\mathrm{gw} = 0.1$.

As has been previously shown in the circular case, we find that different PN approximants
lead to different levels of agreement with NR~\cite{Boyle:2007ft}.
We show here that an
eccentric PN model expanded in terms of the mean motion $n = 2 \pi
/P$, where $P$ is the orbital period, leads to significant
disagreements with NR, whereas using the frequency-related variable $x
= \left ((2 \pi + \Delta \phi)/P\right)^{2/3}$, where $\Delta \phi$ is
the precession angle per period, gives much better agreement.

In Section \ref{sec:pnmodel}, we describe the eccentric PN model we
will use in our comparisons.  The PN expressions are given in outline
form to make clear precisely how we are constructing the solutions;
the full expressions are given in the appendix.  In Section
\ref{sec:nrsims}, we describe the methods used in our NR simulations
that have not previously been described; specifically, we discuss the
method of constructing initial data parameters with eccentricity $e
\approx 0.1$.  We present the results of the numerical simulations in
Section \ref{sec:nrresults}, along with an analysis of the errors.  In
Section \ref{sec:pnmodelfitting}, we describe the method we use for
matching NR and PN waveforms.  Section \ref{sec:eccpncompare} contains
the main result of this paper, which is the comparison of the PN and
NR solutions. Finally, we discuss the consequences of the results and
our plans for future work in Section \ref{sec:conclusions}.

\section{Methods}
\label{sec:methods}

\subsection{Eccentric post-Newtonian model}
\label{sec:pnmodel}

We first review the solution of eccentric Newtonian orbits, in order
to fix notation and to illustrate our general method for solving the
PN system.  For a detailed treatment, see for example Ref.~\cite{Goldstein2002} \footnote{Note that in Ref.~\cite{Goldstein2002}, the eccentric anomaly is called $\psi$ rather than $u$, and the period $T$ rather than $P$.  The notation used in this work reflects that commonly used
in the PN literature.}.  The system under consideration consists of two point
particles of masses $m_1$ and $m_2$.  The total mass is $M = m_1 +
m_2$.  We will use $M$ as the mass scale for all numerical quantities
in our NR simulations, and work in units in which $G = c = 1$.  The
reduced mass is $\mu = m_1 m_2 / M$ and the symmetric mass
ratio is $\eta = \mu / M$. We will give expressions for arbitrary mass
ratios $\eta$, although in this work we will only be considering
equal-mass systems, for which $m_1 = m_2 = M/2$, $\mu = M/4$ and $\eta
= 1/4$.  For Newtonian orbits, the energy $E$ and angular momentum $J$
are constants of the motion and can be expressed in terms of the {\em
  mean motion} $n$ and the eccentricity $e$.  The conservation of $J$
means that the orbit is restricted to a plane.  The mean motion is
related to the orbital (pericenter to pericenter) period $P$ and the
semimajor axis $a$ by $n = 2 \pi / P = a ^{-3/2} M^{1/2}$.  In the Newtonian
case, the pericenter occurs at the same value of the relative angular
coordinate $\phi$ on each orbit; i.e.~there is no precession.  There
is no closed form solution for the relative orbital radius $r$ or
angular frequency $\dot \phi$ in terms of time, but they can be
expressed in terms of the {\em eccentric anomaly} $u$,
\begin{eqnarray}
r &=& a \left [ 1-e \cos u \right ] \\
\dot \phi &=& \frac{n \sqrt{1-e^2}}{\left [ 1-e \cos u \right ] ^2} \, .
\end{eqnarray}
The eccentric anomaly $u$ satisfies Kepler's equation,
\begin{eqnarray}
l &=& u - e \sin u \, ,\label{eqn:kepler}
\end{eqnarray}
where the {\em mean anomaly} $l$ is given by $\dot l = n$.  Since $n$
is a constant, we can integrate to obtain $l = n(t-t_0)$ and
Eq.~(\ref{eqn:kepler}) is a transcendental algebraic equation for $u$,
which can be solved numerically, for example by Newton's method, at
each $t$.  Thus we can obtain $r$ and $\dot \phi$ (and hence $\dot r$
and $\phi$) at any time $t$.  Each orbit is parametrized by the
constants $n$, $e$, $\phi_0 \equiv \phi(t_0)$ and $l_0 \equiv l(t_0)$.

The Newtonian system is conservative in the sense that it admits a
conserved energy and angular momentum, which can be expressed in terms
of the constants $n$ and $e$.  One can also derive conservative
equations in the PN case; the Newtonian equations for $r$, $\dot \phi$
and $l$ are modified by the addition of higher order (in $n$) terms.
In the PN case, the quasi-Keplerian parametrization leads to three
eccentricities, $e_t$, $e_r$ and $e_\phi$, representing deviations
from circular motion in $t$, $r$ and $\phi$, but these are related to
each other by PN equations and it is sufficient to consider just
$e_t$.  To Newtonian order, all three are equal.

In the conservative PN equations, $n$ and $e_t$ are still constants,
but the orbits precess.  Note that the period $P$ of the orbit is
defined to be the time from pericenter to pericenter, and due to the
effects of precession, this is {\em not} the time to go from angular
coordinate $\phi$ to $\phi + 2\pi$.  The angle of precession of the
pericenter during one (pericenter to pericenter) orbit of period $P$
is denoted $\Delta \phi$. Following
Refs.~\cite{Arun:2007rg,Arun:2007sg}, we define
\begin{align}
\omega \equiv& \frac{2 \pi + \Delta \phi}{P}
\end{align}
to be the angle swept out by the orbit from pericenter to pericenter
in one period $P$.  Note that in the conservative PN system, this is a
constant.  In the circular case, where $\dot \phi$ is a constant, we
have $\omega = \dot \phi$.  We will investigate two different PN
models which differ only in the choice of variable used (and hence in
higher order uncontrolled remainder terms).  In Ref.~\cite{2006PhRvD..73l4012K},
the eccentric system is described in terms of the mean motion $n$ and
the eccentricity $e_t$.  We present the equations here in terms of the
variable $x \equiv \left ( M \omega \right ) ^{2/3}$ and $e_t$.  We
call the two resulting PN models the $x$-model and the $n$-model.  See
Sec.~\ref{sec:whynotn} for more discussion of these two models.

We now give the 3~PN conservative orbital dynamics; we work throughout
in modified harmonic coordinates, in which these expressions have been
derived~\cite{Memmesheimer:2004cv}.  The 3~PN conservative dynamics were
first determined in Ref.~\cite{Memmesheimer:2004cv}, and were written
out explicitly in terms of $n$ and $e_t$ in
Ref.~\cite{2006PhRvD..73l4012K}. Here, for brevity, we will omit
lengthy high order PN expansions; the full expressions are available
in the appendix.  The abbreviated forms of the conservative dynamics,
in terms of $x$ and $e_t$, are
\begin{align}
r/M &= \left [ 1-e_t \cos u \right ] x^{-1} + r_\mathrm{1PN} + r_\mathrm{2PN} x \nonumber \\
&\quad + r_\mathrm{3PN} x^2 + \mathcal{O}(x^3) \label{eqn:rpn}\\
M \dot \phi &= \frac{1 \sqrt{1-e_t^2}}{\left [ 1-e_t \cos u \right ] ^2} x^{3/2} + \dot \phi_\mathrm{1PN} x^{5/2} + \dot \phi_\mathrm{2PN} x^{7/2} \nonumber \\
&\quad +  \dot \phi_\mathrm{3PN} x^{9/2}  + \mathcal{O}(x^{11/2}) \label{eqn:phidotpn} \\
l &=  u - e_t \sin u + l_\mathrm{2PN} x^2 \nonumber \\
&\quad + l_\mathrm{3PN} x^3  + \mathcal{O}(x^4) \label{eqn:keplerpn} \\
M \dot l &= M n = x^{3/2} + n_\mathrm{1PN}x^{5/2} + n_\mathrm{2PN} x^{7/2} \nonumber \\
&\quad + n_\mathrm{3PN} x^{9/2} + \mathcal{O}(x^{11/2}) \label{eqn:ldotpn} \, ,
\end{align}
where the quantities $r_\mathrm{1PN}, \dot \phi_\mathrm{1PN}, \ldots$
are functions of $e_t$ and $u$, but $n_\mathrm{1PN}, \ldots$ are
functions only of $e_t$.  Since the right hand side of
Eq.~(\ref{eqn:ldotpn}) is given in terms of the constants $x$ and
$e_t$, it can be trivially integrated to give $l(t)$ in terms of an
integration constant $l_0$ at some $t_0$ .  So given constants $x$,
$e_t$ and $l_0$, we can solve Eq.~(\ref{eqn:keplerpn}) numerically for
$u$ at each $t$ by root-finding, then insert $u$ into
Eqs.(~\ref{eqn:rpn}--\ref{eqn:phidotpn}) to obtain the coordinate motion
of the conservative 3~PN system.

The conservative system is expected to be a good approximation on
timescales over which the energy and angular momentum lost to
gravitational radiation is negligible.  To go beyond this
approximation, we will model these losses adiabatically; i.e.~they
will be averaged over the orbital period.  The losses are derived by
computing the gravitational wave energy and angular momentum flux at
infinity and equating the energy and angular momentum radiated to that
lost from the system.  The equations for $\dot E$ and $\dot J$ can be
used to derive equations for $\dot x$ and $\dot e_t$.  The equations
to 2~PN order are given in Ref.~\cite{2006PhRvD..73l4012K} in terms of
$n$ and $e_t$.  In terms of $x$ and $e_t$, we have
\begin{align}
M \dot x &= \frac{2 \eta }{15(1-e_t^2)^{7/2}} \left ( 96 + 292 e_t^2 + 37 e_t^4
\right ) x^5 + \dot x_\mathrm{1PN} x^{6} \nonumber \\
&\quad  + \dot x_\mathrm{1.5PN} x^{13/2} + \dot x_\mathrm{2PN} x^{7} + \mathcal{O}(x^{15/2}) \\
M \dot e &=  \frac{-e  \eta }{15 (1-e_t^2)^{5/2}} \left ( 304 + 121 e_t^2 \right ) x^4 + \dot e_\mathrm{1PN} x^{5} \nonumber \\
&\quad  + \dot e_\mathrm{1.5PN} x^{11/2} + \dot e_\mathrm{2PN} x^{6} + \mathcal{O}(x^{13/2}) \, .
\end{align}
Since the evolution is treated adiabatically, the functions $\dot
x_\mathrm{1PN}, \dot e_\mathrm{1PN}, \ldots$ depend only on $e_t$, and
not on $u$.  Hence, the adiabatic evolution equations for $x$ and
$e_t$ form a closed system, and can be solved independently of the
Kepler equation.  Given initial conditions $x(0)$ and $e_{t}(0)$, we
can solve the system of ODEs numerically to obtain $x(t)$ and
$e_t(t)$.

In the presence of time-varying $x$ and $e_t$, Eq.~(\ref{eqn:ldotpn})
must be integrated to obtain $l(t)$.  The rest of the computation
proceeds as in the conservative case; $u$ is determined numerically by
root-finding in Eq.~(\ref{eqn:keplerpn}), and then $u$, $x$ and $e_t$
are inserted at each time into Eqs.(~\ref{eqn:rpn}--\ref{eqn:phidotpn}).

We use analytical expressions for the functions $r_\mathrm{1PN}, \dot
\phi_\mathrm{1PN}, \ldots$ to obtain numerical solutions for $r$ and
$\dot \phi$, but due to the complexity of the expressions for $\dot r$
and $\phi$, we choose to obtain $\dot r$ and $\phi$ by numerically
differentiating and integrating the $r$ and $\dot \phi$ solutions
respectively.  This makes a difference to terms at higher PN orders
that we are currently neglecting.

We have checked our expressions for $r$ and $\dot \phi$, as well as
the 3~PN Kepler equation, by deriving them from the orbital
elements in terms of $E$ and $h$~\cite{Memmesheimer:2004cv}, and
comparing with the explicit expressions
in terms of $n$ and $e_t$~\cite{2006PhRvD..73l4012K}.
This completes the description of the coordinate motion.

Since the NR and PN solutions are in different coordinate systems, we
must compare them using some coordinate-independent quantity.  We will
use the gravitational wave frequency; specifically the $\ell = 2$, $m
= 2$ mode of the Newman-Penrose $\Psi_4$ quantity, as it is readily
available from the NR simulations.

The complex PN waveform strain is given (to leading Newtonian order)
by
\begin{align}
h &= h_+ - i h_\times \label{eqn:hincoords1}\\
h_+ &= - \frac{M\eta}{R}  \left \{\left(\cos ^2\theta +1\right)
  \bigg[\cos 2 \phi'
    \left(-\dot{r}^2+r^2 \dot{\phi
   }^2+\frac{M}{r}\right) \right.  \nonumber\\
& \quad  +2 r \dot{r} \dot{\phi } \sin 2 \phi' \bigg ] 
\left. +\left(-\dot{r}^2-r^2 \dot{\phi }^2+\frac{M}{r}\right) \sin ^2 \theta \right \} \\
h_\times &= -\frac{2 M \eta}{R}  \cos \theta  \bigg\{\left(-\dot{r}^2+r^2
   \dot{\phi }^2+\frac{M}{r}\right) \sin 2 \phi' \nonumber \\
& \quad -2 r \cos 2 \phi' \dot{r} \dot{\phi }\bigg\} \, \label{eqn:hincoords2},
\end{align}
where $\phi' \equiv \phi - \varphi$, and $\theta$ and $\varphi$ are
the spherical polar angles of the observer.
Eqs.~\ref{eqn:hincoords1}--\ref{eqn:hincoords2} are taken from
Ref.~\cite{Gopakumar:2001dy} but with the sign convention for the $\cos 2\phi'$
and $\sin 2\phi'$ terms of Refs.~\cite{2006PhRvD..73l4012K,
  Damour:2004bz}.
Using only the leading
(quadrupolar) contribution to $h$ is called the {\em restricted}
waveform approximation.  
The strain $h$ can be decomposed into
spin-weight $s = -2$ spherical harmonics, and the $\ell = 2, m = 2$
mode is given by
\begin{align}
h^{22} &= \int {}_{-2} {Y^2_2}^*(\theta,\varphi) h(\theta,\varphi) d\Omega \\
&= - \frac{4 M \eta e^{ -2i \phi}}{R} \sqrt{\frac{\pi}{5}} \left ( \frac{M}{r} + (\dot \phi r + i \dot r)^2 \right ) \, , \label{eqn:h22}
\end{align}
where ${}_{-2} Y^2_2(\theta,\varphi) = \frac{1}{2} e^{2 i \varphi
}\sqrt{5/\pi} \cos ^4\left(\theta/2\right)$.  We insert the
coordinates, $\phi$, $\dot \phi$, $r$, $\dot r$, into
Eq.~(\ref{eqn:h22}) to obtain the $\ell = 2, m = 2$ spin-weight $s =
-2$ \cite{Goldberg:1966uu} mode of the waveform strain.  Finally,
using $\Psi_4^{22} = \ddot h^{22}$ we differentiate $h^{22}$ twice
with respect to time to obtain the (complex) $\ell = 2, m = 2$ mode of
$\Psi_4$.  This is split into amplitude and phase, and undetermined
additive multiples of $2\pi$ in the phase are determined by
continuity.

We have described one procedure for constructing PN eccentric
waveforms.  Note that this is not unique; different procedures will
differ by the `uncontrolled remainder terms' of higher PN order than
we have considered.  Specifically, we have chosen to solve the 2~PN
truncated adiabatic evolution equations for $x$ and $e_t$ numerically,
rather than constructing an analytic expansion for the solution and
then truncating it to 2~PN.  This makes a difference to the solution
in the circular case \cite{2007PhRvD..75l4018B}, and has been
shown \cite{Boyle:2007ft} to give better phase agreement with NR.
In Ref.~\cite{Boyle:2007ft}, the circular
waveform constructed using this approach is named TaylorT4, and the waveform phase agrees
significantly better with NR than the TaylorT1, TaylorT2 and TaylorT3
approximants.  For simplicity, we have also limited the computation of
the waveform as a function of the coordinates to Newtonian
(quadrupolar) accuracy, and restrict our comparisons to the phase
rather than the amplitude.  Higher order corrections to the waveforms
are available~\cite{Will:1996zj,Gopakumar:1997bs}, though not in a
form which is convenient to use in this work.  We have also
chosen to construct some derivative quantities by numerical
differentiation; where this is the case, we have verified that the
effects of discretization on the resulting waveform phase are much
smaller than any numerical errors we have in our NR simulations.

\subsection{Numerical relativity methods}
\label{sec:nrsims}

Our NR simulations are based on the {\em moving punctures} approach
without excision~\cite{Baker:2005vv,Campanelli:2005dd}.  Initial data
representing the binary black hole system is constructed with a
conformally flat metric and Bowen-York extrinsic curvature, and the
constraints are solved using the
\texttt{TwoPunctures}~\cite{Ansorg:2004ds} spectral code.  The
evolution in time is performed using our BSSN~\cite{Nakamura87,
Shibata95, baumgarte_shapiro} finite differencing code generated using
the \texttt{Kranc}~\cite{Husa:2004ip} code generation package.  The
\texttt{Cactus}~\cite{cactusweb1} infrastructure is used for
parallelization, I/O and parameter handling, and for adaptive mesh
refinement we use \texttt{Carpet}~\cite{Schnetter-etal-03b}.  The code
has been previously described in more detail~\cite{vaishnav-2007},
however we have since modified it to use sixth order spatial finite
differencing as described in Ref.~\cite{husa-2007} in order to improve
accuracy.  We here use 9 levels of box-in-box mesh refinement, where
the outermost (base) grid covers the domain $x^i \in [-384, 384]$.  On
the outer boundary, a simple spherical outgoing wave boundary
condition is applied to each variable as is conventional for finite
differencing BSSN codes in NR (see Ref.~\cite{Alcubierre02a} for more
details).  Formally, this boundary condition respects neither the
constraints nor the characteristic structure of the equations.  For
very short simulations, it is possible to place the outer boundary far
enough out that it is causally disconnected from the coordinate
spheres on which the waveforms are computed, but for the long
simulation we present here it is not computationally feasible to do
this in our code.  A discussion of the possible errors introduced can be found in
Sec.~\ref{sec:nrresults}.

The free parameters in the Bowen-York extrinsic curvature are the
coordinate locations and linear momenta of the two black holes.  We
obtain these parameters using the conservative 3~PN expressions for
eccentric orbits~\cite{2006PhRvD..73l4012K}.  These
expressions require specification of the two constants, $e_t$ and $n$
(the eccentricity and
mean motion).  We choose $n = 0.0156/M$ and $e_t = 0.1$, and
compute the coordinate separation, $r$, from the 3~PN expression in
terms of $n$ and $e_t$, and use it in the Bowen-York extrinsic
curvature.  The tangential linear momentum of each black hole at
apocenter, $p_y$, is obtained from $J = p_y r$, where $J$ is computed
as a PN expansion in $n$ and $e$.  We solve iteratively for the base
mass parameters to ensure that the irreducible masses of the black
holes are $m_1 = m_2 = 0.5 M$, where $M$ is a mass scale.  As such,
the mass scale $M$ is the sum of the irreducible masses of the black
holes.  This procedure results in initial coordinate locations
$x^i_\pm = ( \pm 7.1570737463, 0, 0) M$, initial linear momenta $P^i_\pm =
(0, \pm 0.07191137095, 0) M$, and initial bare masses
$m_\mathrm{bare}^\pm = 0.4903157830 M$.  The resulting spacetime has
ADM mass $M_\mathrm{ADM} = 0.991413 M$.

This choice of initial data parameters has the following
limitations.  Firstly, only the conservative PN expressions have been
used, which means that there is no consideration of the inspiral
velocity.  Secondly, there will be an error in the parameters due to
the truncation of the PN series.  Thirdly, the use of PN parameters
(in this case in harmonic coordinates) directly substituted into the
Bowen-York extrinsic curvature, assumes that the differences in the
coordinate systems are small.  We will see later that these initial
data parameters agree reasonably well with the subsequent evolution.

\subsection{Fitting the post-Newtonian model to numerical relativity data}
\label{sec:pnmodelfitting}

We now discuss our method for determining a PN model which corresponds
to our numerical simulation results.  The PN approximation is very
accurate when the binary system is far separated, becomes less
accurate in the later stages of inspiral, and is no longer valid
during some period leading up to merger.  Using NR, we can simulate
the late inspiral and merger.  Ultimately, we would like to construct
a waveform which most closely resembles one that would be observed in
nature from early inspiral all the way through to merger.  We will
assume that the NR result gives the final part of this hypothetical
{\em full} waveform, and use a PN waveform to approximate the full
waveform before the start of the NR one. In this paper, we will not
construct a hybrid waveform from the PN and NR results.

In this work, we will look for
agreement in the gravitational wave frequency of the $\ell = 2, m = 2$
mode of $\Psi_4$,
\begin{equation}
\omega_\mathrm{gw} \equiv \dot \phi_\mathrm{gw} = \frac{d}{dt} \arg \Psi_4^{22} \, ,
\end{equation}
as is common in the circular case.  We will use the suffix `gw' to
indicate that the quantity we are considering is related to the
gravitational wave, and not the coordinate motion.
We choose a time interval $[t_1,t_2]$ in the numerical simulation and
use least
squares fitting to determine the parameters of the PN model that best
fits the numerical data in that interval.  We will find in Section
\ref{sec:nrresults} that the black hole masses in the numerical
simulations are essentially constant at $m_1 = m_2 = 0.5 M$ for the
inspiral part of the simulation, so we do not fit for the masses when
matching to PN.  Thus, the eccentric PN model is determined uniquely
by a choice of the functions $X$, $e_t$, $l$ and $\phi$ at a given
time $t_0$, where $X = x$ or $n$ depending on the PN model being
constructed.
We define initial conditions
\begin{equation}
y_0 \equiv \left [ X_0, e_0, l_0, \phi_0 \right ] \, ,
\end{equation}
and the residual
\begin{eqnarray}
Q(y_0) &\equiv& \frac{1}{N}\sum_{t \in I}
\left [ \omega_\mathrm{PN}(t; y_0) - \omega_\mathrm{NR}(t) \right ] ^2 \, ,
\end{eqnarray}
using points $t$ from the numerical simulation in the interval $[t_1,t_2]$.
$Q$ is then minimized numerically over $y_0$, where for each $y_0$,
the PN equations must be solved to construct the waveform.  The
minimization requires an initial estimate of $y_0$.  We find that
using a local minimization method (for example, the principal axis
method) can lead to inconsistent results.  Specifically, the final
fitted parameters show a dependence on the initial estimate due to the
existence of local minima in the residual.  Instead, we use a global
minimization method, requiring an order of magnitude more iterations
(typically around 5000),
and hence increased computational resources.  We find that
minimization by the method of differential evolution, as implemented
in Mathematica's {\em NMinimize} function, works well.  A typical
minimization for a given fitting window takes about 20 minutes on a
laptop.  Note that
since the wave frequency $\omega_\mathrm{gw}$ is independent of
$\phi_0$, in practice we determine $\phi_0$ by a separate least
squares fit between the PN and NR waveform phases.

So, given a fitting interval, we can determine a PN model, identified
by the parameters $y_0$.  If the model and data matched exactly, the
fitted parameters $y_0$ would be independent of the fitting interval.
However, the errors in the PN approximation cause the fit to be
imperfect. If these errors are large, the dependence on the fitting
interval will be significant.

Once the parameters $y_0$ have been obtained, we can use these
parameters
to construct a final PN
model, which will be the model that best approximates the full
solution in the fitting interval.

\section{Results}

\subsection{Numerical relativity simulation results}
\label{sec:nrresults}

In this section, we describe the results of our NR simulations, and
analyze the numerical errors.  The PN model gives the limiting form of
the waveform at large distances from the source, whereas
in the numerical code we compute the waveform on coordinate spheres of
finite radii $r_\mathrm{ext}/M = \{30, 40, \ldots, 150\}$.  We
therefore extrapolate the waveform to infinite radius using the
method described in Ref.~\cite{Boyle:2007ft}.  
To extrapolate the waveform, we first shift the
waveform measured at each extraction radius in time by the estimated
light propagation time to the extraction sphere, given by the
Schwarzschild tortoise coordinate~\cite{Fiske05},
\begin{equation}
r^\star = r_\mathrm{areal} + 2 M_\mathrm{ADM} \log \left (
\frac{r_\mathrm{areal}}{2 M_\mathrm{ADM}} - 1 \right ) \, ,
\end{equation}
where we approximate $r_\mathrm{areal} \approx r + M_\mathrm{ADM}$
(see Ref.~\cite{Boyle:2007ft} for further details; even if this
relation does not hold exactly, the deviation will be included in our
extrapolation error estimates). The amplitude and phase
are then separately extrapolated by a least squares fit to an $n$th
degree polynomial in $1/r$, $f^n(r) \equiv f_\infty + \sum_{i=1}^n
{a_i}/{r^i}$ at each time $t-r^\star$.  We estimate the error in the
$n$th order extrapolation as $e^n \equiv f^{n+1} - f^n$.  We find that
using extraction radii $r_\mathrm{ext} = \{70, 80, \ldots, 150\}$ in
combination with first order extrapolation gives the best results.
Using higher order extrapolation does decrease the error, but the
extrapolant contains more noise.

\begin{figure}
\includegraphics[trim=1.25cm 0 0 0]{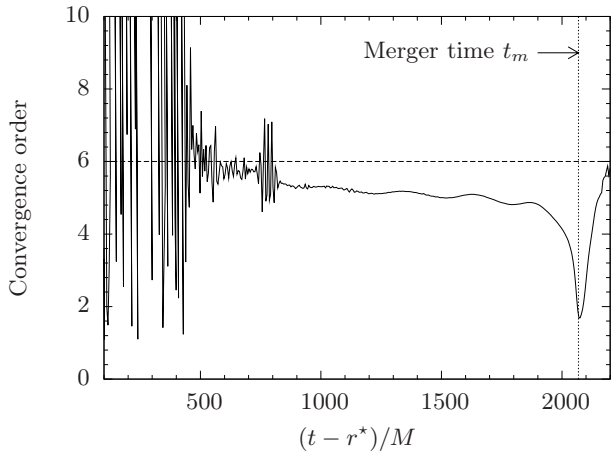}
\caption{Convergence order of the NR gravitational wave phase
$\phi_\mathrm{gw}$.  Deviations from the expected value of 6 may be
caused by lower order components in the code.}
\label{fig:phaseconvorder}
\end{figure}

\begin{figure}
\includegraphics[trim=1cm 0 0 0]{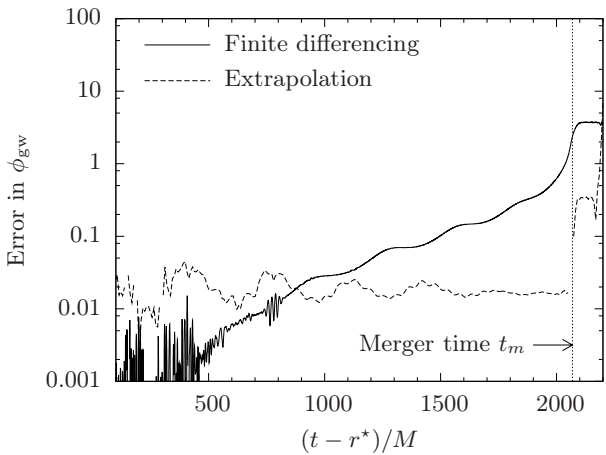}
\caption{Errors in the NR gravitational wave phase $\phi_\mathrm{gw}$
from the effects of finite resolution and extrapolation to infinite
radius}
\label{fig:phaseerrors}
\end{figure}

We ran three simulations at different resolutions in order to assess
the finite differencing error.  The finest refinement boxes were
coordinate cubes of side $1.24 M$ and consisted of $48^3, 64^3, 80^3$
points in the three runs.  This leads to finest grid spacings of
$h_\mathrm{f} = M/38.7, M/51.6, M/64.5$.  To investigate the finite
differencing error, we consider the convergence properties of the
gravitational wave phase,
\begin{equation}
\phi_\mathrm{gw}(t-r^\star) \equiv \arg \left [\Psi_4^{22}(t-r^\star) \right ] \, ,
\end{equation}
extrapolated to infinite radius.  In Fig.~\ref{fig:phaseconvorder}, we
plot the convergence order of $\phi$.  For $t - r^\star < 500$, we see
no clear convergence order, but the differences between the phases at
the three resolutions are less than 0.01 radians.  For $500 < t -
r^\star < 2000$ we see a convergence order which drops from 6 to 5,
after which the order drops to about 1 for a small period around the
merger.  The fact that the convergence order is not clearly 6 may be
explained by the fact that we have second, fourth and sixth order
components in the simulation.  Since we do not have clean sixth order
convergence, we cannot reliably use Richardson extrapolation to obtain
a more accurate result.  However, we can use extrapolation of the
highest two resolutions using the observed approximate convergence
order of 5 to provide an estimate of the error in the solution.  Note
that for the time region we will use for matching with PN ($t <
1000$), the convergence order of 5 is a good approximation.
Figure~\ref{fig:phaseerrors} shows the finite differencing error
estimate compared with the estimate of the error in the extrapolation
to infinite radius.  The dotted line represents the time of the peak
in $|\Psi_4^{22}|$, which is a good indicator of the merger time.
Note the sudden increase in the extrapolation error shortly after the
merger.  Also note that any significant effects arising from numerical
reflections of the waves from refinement boundaries are expected to be
covered by the finite differencing error bars, as these effects should
diminish with increased resolution.

When comparing with PN later, we will add the errors from finite
differencing and from extrapolation in quadrature to provide an
estimate of the overall error in the numerical waveform.  Note that
the approximately exponential growth of the finite differencing error
in Fig.~\ref{fig:phaseerrors} has been previously 
observed in the circular case~\cite{husa-2007}.

\begin{figure}
\includegraphics{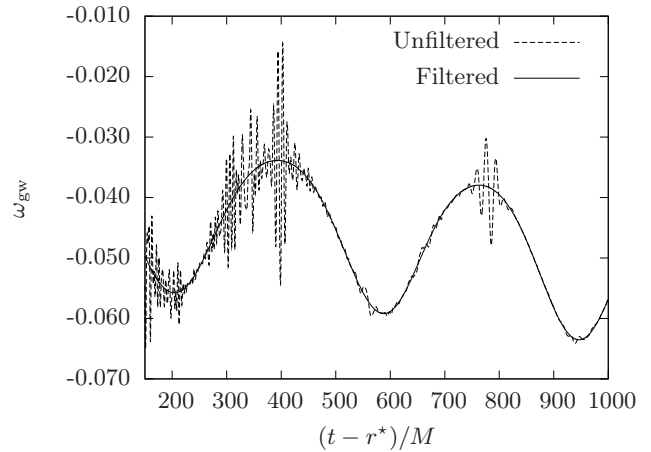}
\caption{Filtering of NR gravitational wave frequency in the Fourier
domain.  The solution in the region containing noise is truncated to
the lowest 30 Fourier modes.}
\label{fig:filtering}
\end{figure}

When comparing NR and PN models, we wish to use the gravitational wave
frequency $\omega_\mathrm{gw}$.  However, as seen before for both
finite differencing~\cite{Hannam:2007ik} and
pseudo-spectral~\cite{Boyle:2007ft} codes in the circular case,
$\omega_\mathrm{gw}$ has noticeable high frequency error at early
times when the amplitude of the radiation is low.  In our case, this
comes from numerical reflections of the initial spurious radiation,
present in the initial data, from mesh refinement boundaries.  Since
this is precisely the regime in which we would like to match with PN,
this high frequency error must be removed.  We find that this can be
achieved very effectively by filtering the noisy region of
$\omega_\mathrm{gw}$ in the Fourier domain.  We first tried using a
moving averages filter, but this tended to systematically reduce the
amplitude of the oscillations in $\omega_\mathrm{gw}$, which we found
unacceptable.  We also chose not to fit a polynomial to
$\omega_\mathrm{gw}$ as has been done in the circular
case~\cite{Hannam:2007ik}, due to the naturally oscillatory nature of
the eccentric signal.  To perform the filtering, we proceeded as
follows.  We first chose an interval of time, $[t_1,t_2]$, in which to
perform the filtering.  We chose $[t_1,t_2] = [80 M, 1680 M]$,
excluding the late inspiral and merger as well as the initial spurious
radiation from the filtering region. We then performed a discrete
Fourier transform of the data, removed all but the lowest 30 modes,
and then inverse transformed.  We found that 30 modes were sufficient
to represent the signal; this was judged by subtracting the filtered
from the unfiltered signal, and observing essentially only noise.
Taking only the first 30 modes corresponds to a frequency cutoff of $
\omega_\mathrm{max} = 30 \times 2 \pi / (t_2-t_1) \approx 0.1 M^{-1}$,
or modes with a period of $T_\mathrm{min} \approx 50 M$.  Note that
this is not comparable to filtering the evolved variables or even
$\Psi^4_{2,2}$; it is the {\em frequency} of $\Psi^4_{2,2}$ that is
being filtered.  Since the original signal is not periodic, Gibbs
phenomena were observed as oscillations near the endpoints of the
filtered region.  We therefore removed a segment of length $80M$ from
the beginning and end of the filtered region before re-inserting the
filtered region into the full signal.  Fig.~\ref{fig:filtering} shows
the result of the filtering.

We have monitored the irreducible masses $M_\mathrm{irr}$ of the
apparent horizons in the lowest resolution simulation.  The computed
mass of each black hole drops from its initial value of 0.5 by only $2
\times 10^{-4} M$ by the time of the merger, and we ascribe this
effect to finite differencing error.  We have not computed horizons at
higher resolutions due to computational expense.  Thus, within our
numerical errors, we do not detect any physical growth of the horizons
during the inspiral, which potentially could have occurred due to
absorption of gravitational wave energy in the initial part of the
simulation, as has been studied in detail in previous
work~\cite{Bode:2007dv}.

The spins of the black holes, as measured using an approximate
technique derived from the isolated horizon formalism~
\cite{2007arXiv0706.2541H,Ashtekar:2004cn}, increase during the
simulation to only $S^z = 10^{-4} /M^2$ before the merger.  This is
independent of finite differencing resolution, but we expect this tiny
spin to be of little consequence to the PN comparison, which does not
contain the effects of spin.

The outer boundary in the simulations is at $x^i = \pm 384 M$, and as
mentioned in Sec.\ref{sec:nrsims}, the boundary condition is a source
of error in the simulation.  To measure the effect of this error, we
have repeated the low resolution simulation, which has only modest
computational cost, with the outer boundary moved to $x^i = \pm 768 M$
by enlarging the coarsest grid.  We find that the effect on the
waveform phase is much smaller than the estimated errors in the high
resolution simulation due to finite differencing and extrapolation to
infinite radius, and we conclude that the outer boundary is not a
significant source of error in the simulation.  In future, with more
accurate simulations, this will need to be addressed further.

The simulations at the three different resolutions consumed
approximately 5000, 11000 and 16000 CPU hours respectively, each one
running on 32 cores of the LoneStar supercomputer.

\subsection{Comparing numerical relativity simulations with post-Newtonian models}
\label{sec:eccpncompare}

We now discuss the results of applying the fitting procedure described
in Section \ref{sec:pnmodelfitting} to the numerical simulation
results.  

\begin{figure}
\includegraphics[trim=0 0 0 -0.5cm]{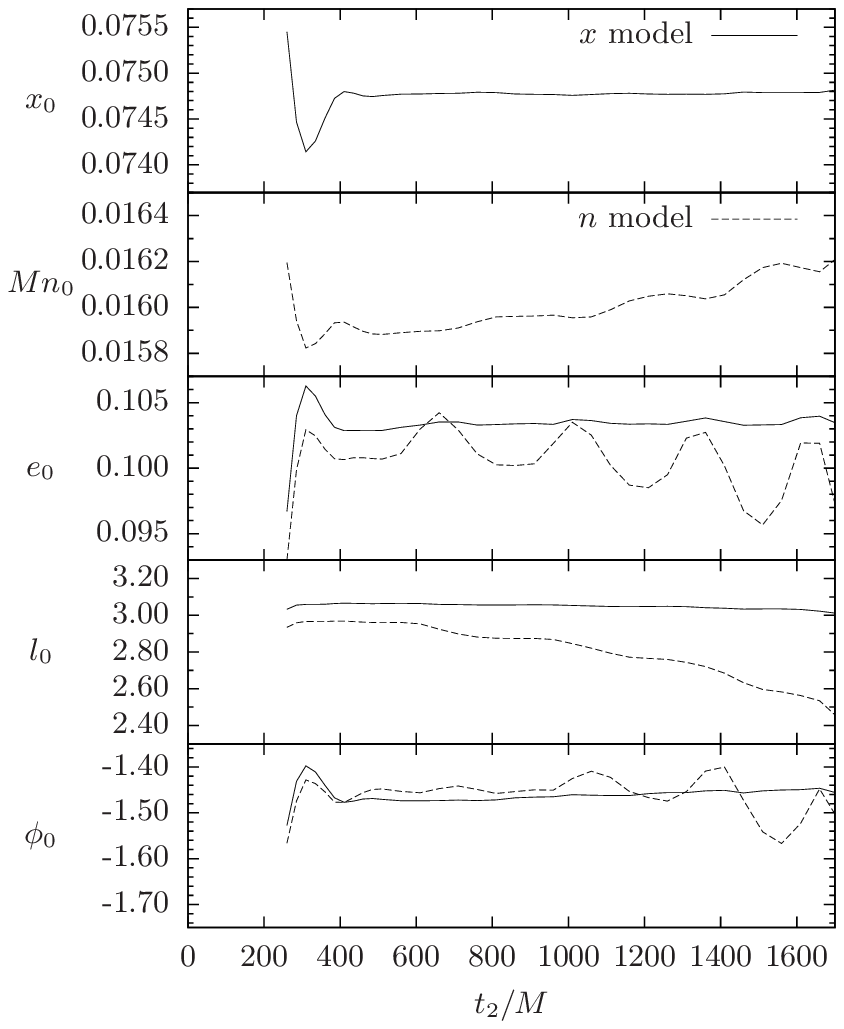}
\caption{PN parameters for the $x$ and $n$ models as determined from
  fitting windows $[t_1, t_2]$ for $t_1 = 210 M$ and various values of
  $t_2$.}
\label{fig:fit-t2}
\end{figure}

\begin{figure}
\includegraphics[trim=0 0 0 -0.5cm]{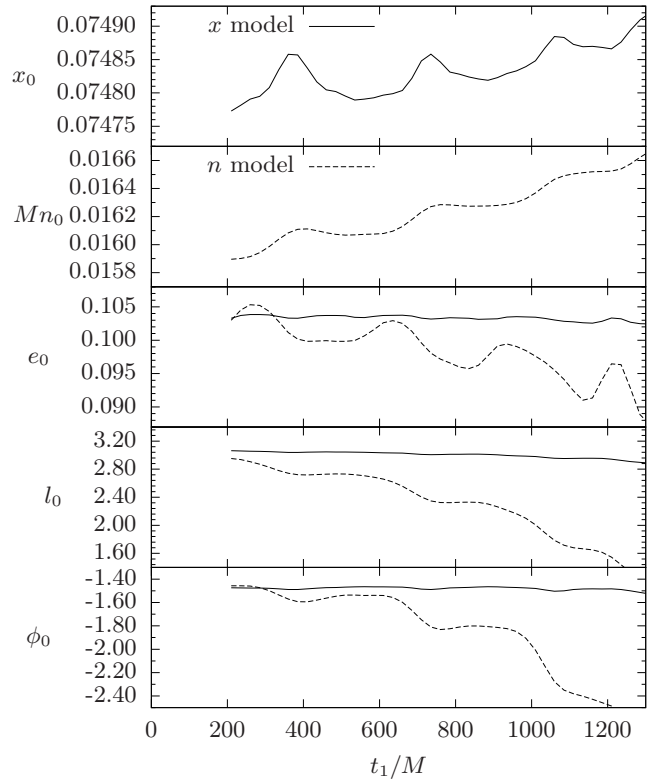}
\caption{PN parameters for the $x$ and $n$ models as determined from
  fitting windows $[t_1, t_2]$ for various values of
  $t_1$ and $t_2 = t_1 + 400 M$.}
\label{fig:fit-t1}
\end{figure}

Figures \ref{fig:fit-t2} and \ref{fig:fit-t1} show the parameters
$\left [x_0, e_0, l_0, \phi_0 \right ]$ (for the x model) and $\left
  [n_0, e_0, l_0, \phi_0 \right ]$ (for the n model) determined by
fits of the NR data to the PN model in fitting intervals $I = [t_1,
t_2]$.
These parameters are the values of the functions $x$, $n$, $e$, $l$ and
$\phi$ at $t-r^\star = 0$.
In Fig.~\ref{fig:fit-t2}, $t_1$ has been kept fixed to a value
at the start of the usable waveform and $t_2$ has been varied.  We see
that the parameters obtained from fits using the $x$ model vary much
less with the fitting window length than those using the $n$ model.
Specifically, we see that for both models the fitted parameters
oscillate significantly for intervals of less than $\sim 400 M$, but
for the $x$ model these variations die away as the interval is
increased beyond this.  From the initial data parameters, the orbital
period is $P = 403 M$.  It may be that over timescales smaller then
the orbital period, there are unmodeled non-adiabatic oscillations in
the NR result which are averaged out when larger fitting intervals are
used.  These oscillations may cause the fit to become worse for small
intervals.  For the $n$ model we see strong oscillations of a period
$\sim 400 M$ roughly corresponding to the period of the oscillations
in $\omega_\mathrm{gw}$ itself.  In order to determine the effect on
the parameters of the interval location, we choose an interval width
of $400 M$ and vary $t_1$ in Fig.~\ref{fig:fit-t1}.  Here again we see
that the $x$ model shows much more consistent behavior than the $n$
model.

In order to choose a unique set of PN parameters, we choose the
earliest possible fitting interval, and take the size of the interval
to approximately correspond to the initial orbital period, $\sim 400
M$, giving a fitting interval $t/M \in [210, 610]$.  The parameters
for this fitting interval are given in Table~\ref{tab:finalparams}.
It is interesting to compare these parameters with the approximate
parameters used to construct the Bowen-York initial data; these are
also given in the table. $x_0$ and $n_0$ agree to within 1\% and 2\%
respectively with the initial data values.  $e_0$ agrees within 0.3\%
between the two PN models, and to 3\% with the initial data value.
$l_0$ agrees to within 0.1 radians between the two models and the
initial data value.  $\phi_0$ agrees to within 0.02 radians between
the two models, but is of the order of $\pi/2$ different from the
initial data value.  This large discrepancy is probably related to the
adjustment of the coordinate system that happens at the start of the
numerical simulation.  Recall that the method for constructing the
initial data parameters was approximate, due to the different
coordinate systems used, so perfect agreement is not expected.

\begin{table}
  \begin{center}
\begin{tabular}{l|l|l|l}
\hline
\hline
Parameter   & $x$-model fit & $n$-model fit & Initial data value \\
\hline 
$x_0$    & $0.0747729$ & -           & $0.0740853$    \\
$n_0$    & -           & $0.0158959$ & $0.0156$       \\
$e_0$    & $0.103291$  & $0.10299$   & $0.1$          \\
$l_0$    & $3.06358$   & $2.9529$    & $\pi = 3.1416$ \\
$\phi_0$ & $-1.47386$  & $-1.45652$  & $0$            \\
\hline
\hline
\end{tabular}
  \end{center}
  \caption{
    Eccentric PN ($x$-model and $n$-model) parameters computed by
    fitting in an interval $[210, 610]$ as well as the parameters
    estimated from the initial data.
    The parameters correspond to the values of the functions $x$,
    $n$, $e$, $l$ and $\phi$ at $t-r^\star = 0$.
    Note that the agreement is not expected to be exact.}
\label{tab:finalparams}
\end{table}

\begin{figure}
\includegraphics[trim=0 0 0 -0.5cm]{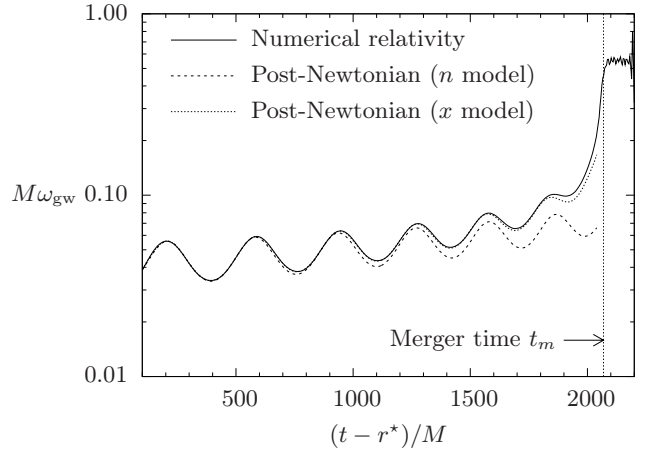}
\caption{Gravitational wave frequency as a function of time from the
NR simulation and two PN models.  The PN $x$-model agrees very well up
to $\approx 1800 M$, whereas the agreement with the $n$-model is
significantly worse.}
\label{fig:extrapom}
\end{figure}

Now that we have estimated the PN model which matches the NR solution
in the fitting interval, we can compare the PN waveform for the $x$
and $n$ models with the NR result.  In Fig.~\ref{fig:extrapom}, we
plot the PN and NR gravitational wave frequencies $\omega_\mathrm{gw}$
and see that there is good agreement with the $x$ model from the start
of the simulation to $t \approx 1800 M$.  That there is such a high
level of agreement with a model which contains so much structure is a
strong validation of both the PN model and the NR simulation.  We also
see on the same plot the much worse agreement obtained using the
$n$-model.

\begin{figure}
\includegraphics[trim=0 0 0 -0.5cm]{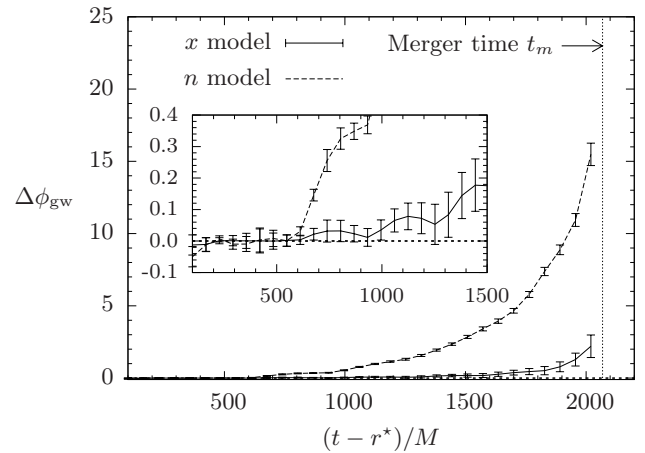}
\caption{Difference in gravitational wave phase between the NR
simulation and the PN $x$-model.  The error bars represent the
estimated errors in the NR simulation.}
\label{fig:extrapphase}
\end{figure}

We now quantify the agreement with the $x$ and $n$-models by considering the
waveform phase differences.  Fig.~\ref{fig:extrapphase} shows the
difference between the NR and PN gravitational wave phases as a
function of $t$.  The error bars represent the uncertainty in the NR
phase from extrapolation to infinite radius and finite differencing
truncation error.  We see that the phase difference
between NR and PN is within 0.1 radians for approximately $1330 M$, or
11 GW cycles.
At $t = 1882 M$, corresponding to $M \omega_\mathrm{gw} = 0.1$, the
phase difference between NR and PN is $\approx 0.7$ radians.

To put the phase difference of 0.7 radians at $M \omega_\mathrm{gw} =
0.1$ into context, we note that the TaylorT4 circular PN model, which
is very similar to our eccentric model with $e = 0$, has been shown to
have a phase difference at $M \omega_\mathrm{gw} = 0.1$ of $\sim 0.3$
radians for 2 PN radiation reaction (see Fig.~22 in
Ref.~\cite{Boyle:2007ft}).  We should be cautious about drawing the
conclusion that the agreement in the circular case is better, however,
as $M \omega_\mathrm{gw} = 0.1$ may not be directly comparable in the
two cases, particularly because $\omega_\mathrm{gw}$ oscillates in the
eccentric case, but is monotonic in the circular case. The steepness
of the phase difference in Fig.~22 in Ref.~\cite{Boyle:2007ft} at that
point makes the comparison very sensitive to the exact point chosen.

\subsection{Choice of post-Newtonian variables}
\label{sec:whynotn}

Throughout this work we have presented the results of fitting two PN
models with NR data.  The two models differ only in the choice of
variable used: the frequency-related variable $x$ or the mean motion
$n$.  Our first attempts at matching the NR simulation with an
eccentric PN model used $n$.  We studied this case extensively, but
found significant disagreement, as has been shown.
Faced with this disagreement,
we studied the (much simpler)
circular case using a simulation~\cite{Hinder:2007qu} with
low-eccentricity initial data~\cite{Husa:2007rh} and a
circular PN model formed by taking our eccentric $n$-model and setting
$e = 0$.  This model is suboptimal as it only has 2~PN radiation reaction, and 3.5 PN
expressions are available for the circular case.  The agreement between NR
and PN is very
poor even in the
circular case using $n$; the gravitational wave phase difference
at $M \omega_\mathrm{gw} = 0.1$ is $\sim 20$ radians.  (Note that one should be
careful about making direct detailed comparisons between the circular
and eccentric cases, due to the ambiguity in the choice of reference
point $M \omega_\mathrm{gw} = 0.1$ due to the eccentric oscillations in $\omega_\mathrm{gw}$.) However, 
expressing the PN equations in
terms of the coordinate angular velocity of the black holes, $\omega$,
as is common in the literature, gives a significant improvement over using $n$; at $M \omega_\mathrm{gw} = 0.1$, the phase difference is 0.8 radians.
This is in broad agreement with the difference of $\sim 0.3$ radians in
Fig.~22 of Ref.~\cite{Boyle:2007ft} for the
TaylorT4 model at 2 PN, accounting for the uncertainty in the choice of comparison time.
This motivated us to
search for a frequency-related variable applicable in the eccentric
case, and we chose to use $x = \left ( M \omega \right )^{2/3}$, for
compatibility with Ref.~\cite{Arun:2007sg} (recall that in the
eccentric case, $\omega \equiv (2\pi+\Delta\phi)/P \ne \dot \phi$), leading to the 
0.7 radian phase difference at $M \omega_\mathrm{gw} = 0.1$ we report here.

\section{Conclusions}
\label{sec:conclusions}

We have presented NR results for an inspiraling eccentric black hole
binary system with initial eccentricity $e \approx 0.1$ and compared them
with two adiabatic eccentric PN models ($x$ and $n$) with 2 PN radiation reaction.
For the $x$ model, the gravitational wave phase
agrees to within $\pm 0.1$ radians between 21 and 11 cycles before
merger.  The difference grows to 0.7 radians at $\approx 5$ cycles before
merger ($M \omega_\mathrm{gw} = 0.1$), in broad agreement with the circular
case at 2 PN order.  One cycle before the merger, the solution to the
PN ODEs diverges, indicating a breakdown of the model.

We found that it was necessary to express the PN model in terms of the
frequency-related variable $x$ rather than the mean motion $n$ to get
this level of agreement.  We conjecture that, when expressed in terms
of $n$, certain higher order PN terms are non-negligible, whereas when
expressed in terms of $x$, they are small, leading to a smaller error
in the PN solution.  This can be likened to 
studies~\cite{Hannam:2007ik,Boyle:2007ft} where different
circular PN approximants of the same order have been shown to have different
errors in the NR regime.  In particular, the TaylorT4
circular model showed
a remarkable agreement in the waveform phase, but there was a noticeable
disagreement in the energy flux~\cite{Boyle:2008ge}.  It has also been shown 
that this remarkable agreement is lost
when spinning systems are considered~\cite{Hannam:2007ik}.  
Our eccentric PN model based
on $x$ is very similar to TaylorT4 as $e \to 0$, so we would expect
the same conclusions to apply.  

Now that it is possible to match NR and PN eccentric waveforms, we
plan to start to construct hybrid templates and begin to assess the
implications for the interferometric detection of gravitational wave
signals from eccentric binaries close to and including merger.
Since 
complete 3~PN radiation reaction terms for the angular momentum flux
have now also been computed, we will be able to compare with a fully 3~PN
model, and expect the agreement with NR to get better closer to the
merger.

\begin{acknowledgments}
  This work was supported in part by NSF grants PHY-0925345 to DS,
  PHY-0653303, PHY-0555436, PHY-0855892, PHY-0914553 to PL,
  PHY-0941417, PHY-0903973 to PL and DS, and PHY-0114375 (CGWP).
  Computations were performed at NCSA and TACC under allocation
  TG-PHY060013N.  Computer algebra and data analysis were performed
  using Mathematica.  The authors thank M.~Ansorg, T.~Bode, A.~Knapp,
  and E.~Schnetter for contributions to the computational
  infrastructure and E.~Bentivegna, J.~Read and N.~Yunes for helpful
  discussions.
\end{acknowledgments}
\clearpage
\begin{widetext}

\appendix

\section{PN expressions}
\label{sec:pnexpressions}

We now present, for reference, the full PN expressions used in this
work.  The expressions for the 3~PN conservative dynamics (i.e.~$r$, $\dot
\phi$, $l$, $n$) can be derived in two ways from the existing
literature.  They are given directly in
Ref.~\cite{2006PhRvD..73l4012K} in terms of $n$ and $e_t$, so all that
remains is to express them in terms of $x$ and $e_t$.  Recall that $x$
is defined as $x \equiv \left ( M \omega \right ) ^{2/3}$ where
$\omega \equiv (2 \pi + \Delta \phi)/P$ and $P = 2 \pi/ n$.  In
Ref.~\cite{Memmesheimer:2004cv}, $\Phi$ is used in place of $\Delta
\phi$, where $\Phi = 2\pi + \Delta \phi$.  This reference gives
expressions for $n$, $e_t$ and $\Phi$ in terms of $E$ and $J$; these
can be used to obtain $n$ in terms of $x$ and $e_t$,
{\small
\begin{align}
M n &= x^{3/2} + n_\mathrm{1PN}x^{5/2} + n_\mathrm{2PN} x^{7/2} + n_\mathrm{3PN} x^{9/2} + \mathcal{O}(x^{11/2}) \\
n_\mathrm{1PN} &= 
\frac{3}{e^2-1}
 \\
n_\mathrm{2PN} &= 
\frac{(26 \eta -51) e^2+28 \eta -18}{4 \big(e^2-1\big)^2}
 \\
n_\mathrm{3PN} &= \frac{-1}{128 (1-e^2)^{7/2}} \bigg [ 
(1536 \eta -3840) e^4+(1920-768 \eta ) e^2-768 \eta +\sqrt{1-e^2} \big(\big(1040 \eta ^2-1760 \eta +2496\big) e^4 \nonumber \\
&\quad +\big(5120 \eta ^2+123 \pi ^2 \eta -17856 \eta +8544\big) e^2+896 \eta ^2-14624 \eta +492 \eta  \pi ^2-192\big)+1920
 \bigg] \, ,
\end{align}
}
where, for brevity, we have written $e \equiv e_t$.  This expression
for $n$ is then substituted into the conservative expressions in
Ref.~\cite{2006PhRvD..73l4012K} to obtain the conservative expressions
in terms of $x$ and $e_t$, dropping any resulting terms which are higher than 3 PN.  
Alternatively, we can derive these
expressions by taking the expressions for the orbital elements in
Ref.~\cite{Memmesheimer:2004cv}, along with the expressions for $r$
and $\dot \phi$, all in terms of $E$ and $J$.  By both methods, we
obtain for the separation $r$,
{\small
\begin{align}
r/M &= r_\mathrm{0PN} x^{-1} + r_\mathrm{1PN} + r_\mathrm{2PN} x + r_\mathrm{3PN} x^2 + \mathcal{O}(x^3)\\
 r_\mathrm{0PN} &= 
1-e \cos (u)

 \\
 r_\mathrm{1PN} &= 
\frac{2 (e \cos (u)-1)}{e^2-1}+\frac{1}{6} (2 (\eta -9)+e (7 \eta -6) \cos \
(u))

 \\
 r_\mathrm{2PN} &= \frac{1}{(1-e^2)^2} \Bigg [ 
\frac{1}{72} \left(8 \eta ^2+30 \eta +72\right) e^4+\frac{1}{72} \left(-16 \
\eta ^2-876 \eta +756\right) e^2+\frac{1}{72} \left(8 \eta ^2+198 \eta \
+360\right)

 \nonumber \\
&\quad + 
\left(\frac{1}{72} \left(-35 \eta ^2+231 \eta -72\right) e^5+\frac{1}{72} \
\left(70 \eta ^2-150 \eta -468\right) e^3+\frac{1}{72} \left(-35 \eta ^2+567 \
\eta -648\right) e\right) \cos (u)

 \nonumber \\
&\quad + 
\sqrt{1-e^2} \left(\frac{1}{72} (360-144 \eta ) e^2+\frac{1}{72} (144 \eta \
-360)+\left(\frac{1}{72} (180-72 \eta ) e^3+\frac{1}{72} (72 \eta -180) \
e\right) \cos (u)\right)

 \Bigg ] \\
r_\mathrm{3PN} &= \frac{1}{181440 (1-e^2)^{7/2}}\Bigg [ 
\big(-665280 \eta ^2+1753920 \eta -1814400\big) e^6
+\big(725760 \eta ^2-77490 \pi ^2 \eta +5523840 \eta \nonumber \\
& \quad -3628800\big) e^4 
+\big(544320 \eta ^2+154980 \pi ^2 \eta -14132160 \eta +7257600\big) e^2
-604800 \eta ^2+6854400 \eta \nonumber \\
& \quad +\big(\big(302400 \eta ^2-1254960 \eta +453600\big) e^7+\big(-1542240 \eta ^2-38745 \pi ^2 \eta +6980400 \eta -453600\big) e^5 \nonumber \\
& \quad +\big(2177280 \eta ^2+77490 \pi ^2 \eta -12373200 \eta +4989600\big) e^3+\big(-937440 \eta ^2-38745 \pi ^2 \eta +6647760 \eta \nonumber \\
& \quad -4989600\big) e\big) \cos (u)+\sqrt{1-e^2} \big(\big(-4480 \eta ^3-25200 \eta ^2+22680 \eta -120960\big) e^6+\big(13440 \eta ^3+4404960 \eta ^2 \nonumber \\
& \quad +116235 \pi ^2 \eta -12718296 \eta +5261760\big) e^4+\big(-13440 \eta ^3+2242800 \eta ^2+348705 \pi ^2 \eta -19225080 \eta  \nonumber \\
& \quad + 16148160\big) e^2+4480 \eta ^3+45360 \eta ^2-8600904 \eta +\big(\big(-6860 \eta ^3+550620 \eta ^2-986580 \eta +120960\big) e^7 \nonumber \\
& \quad +\big(20580 \eta ^3-2458260 \eta ^2+3458700 \eta -2358720\big) e^5+\big(-20580 \eta ^3-3539340 \eta ^2-116235 \pi ^2 \eta +20173860 \eta  \nonumber \\
& \quad -16148160\big) e^3+\big(6860 \eta ^3-1220940 \eta ^2-464940 \pi ^2 \eta +17875620 \eta -4717440\big) e\big) \cos (u)+116235 \eta  \pi ^2  \nonumber \\
& \quad +1814400\big)-77490 \eta  \pi ^2-1814400
 \Bigg ] \, .
\end{align}
}
The relative angular velocity $\dot \phi$ is found to be
{ 
\begin{align}
M \dot \phi &=  \dot \phi_\mathrm{0PN} x^{3/2} + \dot \phi_\mathrm{1PN} x^{5/2} + \dot \phi_\mathrm{2PN} x^{7/2} +  \dot \phi_\mathrm{3PN} x^{9/2}  + \mathcal{O}(x^{11/2})  \\
 \dot \phi_\mathrm{0PN} &= 
\frac{\sqrt{1-e^2}}{(e \cos (u)-1)^2}
 \\
 \dot \phi_\mathrm{1PN} &= 
-\frac{e (\eta -4) (e-\cos (u))}{\sqrt{1-e^2} (e \cos (u)-1)^3}
 \\
 \dot \phi_\mathrm{2PN} &= \frac{1}{12 (1-e^2)^{3/2} (e \cos(u)-1)^5} \bigg [ 
\big(-12 \eta ^2-18 \eta \big) e^6+\big(20 \eta ^2-26 \eta -60\big) e^4+\big(-2 \eta ^2+50 \eta +75\big) e^2+\Big[\big(-14 \eta ^2 \nonumber \\
& \quad +8 \eta -147\big) e^5+\big(8 \eta ^2+22 \eta +42\big) e^3\Big] \cos ^3(u)+\Big[\big(17 \eta ^2-17 \eta +48\big) e^6+\big(-4 \eta ^2-38 \eta +153\big) e^4+\big(5 \eta ^2-35 \eta  \nonumber \\
& \quad +114\big) e^2\Big] \cos ^2(u)-36 \eta +\Big[\big(-\eta ^2+97 \eta +12\big) e^5+\big(-16 \eta ^2-74 \eta -81\big) e^3+\big(-\eta ^2+67 \eta -246\big) e\Big] \cos (u) \nonumber \\
& \quad +\sqrt{1-e^2} \Big[e^3 (36 \eta -90) \cos ^3(u)+\big((180-72 \eta ) e^4+(90-36 \eta ) e^2\big) \cos ^2(u)+\big((144 \eta -360) e^3 \nonumber \\
& \quad +(90-36 \eta ) e\big) \cos (u)+e^2 (180-72 \eta )+36 \eta -90\Big]+90
 \bigg ] \\
 \dot \phi_\mathrm{3PN} &= \frac{1}{13440 (1-e^2)^{5/2} (e \cos(u)-1)^7} \bigg [ 
\big(10080 \eta ^3+40320 \eta ^2-15120 \eta \big) e^{10}+\big(-52640 \eta ^3-13440 \eta ^2+483280 \eta \big) e^8 \nonumber \\
& \quad +\big(84000 \eta ^3-190400 \eta ^2-17220 \pi ^2 \eta -50048 \eta -241920\big) e^6+\big(-52640 \eta ^3+516880 \eta ^2+68880 \pi ^2 \eta  \nonumber \\
& \quad -1916048 \eta +262080\big) e^4+\big(4480 \eta ^3-412160 \eta ^2-30135 \pi ^2 \eta +553008 \eta +342720\big) e^2+\big(\big(13440 \eta ^3+94640 \eta ^2 \nonumber \\
& \quad -113680 \eta -221760\big) e^9+\big(-11200 \eta ^3-112000 \eta ^2+12915 \pi ^2 \eta +692928 \eta -194880\big) e^7+\big(4480 \eta ^3+8960 \eta ^2 \nonumber \\
& \quad -43050 \pi ^2 \eta +1127280 \eta -147840\big) e^5\big) \cos ^5(u)+\big(\big(-16240 \eta ^3+12880 \eta ^2+18480 \eta \big) e^{10}+\big(16240 \eta ^3-91840 \eta ^2 \nonumber

\\
& \quad  
+17220 \pi ^2 \eta -652192 \eta +100800\big) e^8+\big(-55440 \eta ^3+34160 \eta ^2-30135 \pi ^2 \eta -2185040 \eta +2493120\big) e^6 \nonumber \\
& \quad +\big(21840 \eta ^3+86800 \eta ^2+163590 \pi ^2 \eta -5713888 \eta +228480\big) e^4\big) \cos ^4(u)+\big(\big(560 \eta ^3-137200 \eta ^2+388640 \eta \nonumber \\
& \quad  +241920\big) e^9+\big(30800 \eta ^3-264880 \eta ^2-68880 \pi ^2 \eta +624128 \eta +766080\big) e^7+\big(66640 \eta ^3+612080 \eta ^2-8610 \pi ^2 \eta  \nonumber \\
& \quad  +6666080 \eta -6652800\big) e^5+\big(-30800 \eta ^3-294000 \eta ^2-223860 \pi ^2 \eta +9386432 \eta \big) e^3\big) \cos ^3(u)+67200 \eta ^2 \nonumber \\
& \quad +\big(\big(4480 \eta ^3-20160 \eta ^2+16800 \eta \big) e^{10}+\big(3920 \eta ^3+475440 \eta ^2-17220 \pi ^2 \eta +831952 \eta -725760\big) e^8+\big(-75600 \eta ^3 \nonumber \\
& \quad  +96880 \eta ^2+154980 \pi ^2 \eta -3249488 \eta -685440\big) e^6+\big(5040 \eta ^3-659120 \eta ^2+25830 \pi ^2 \eta -7356624 \eta +6948480\big) e^4 \nonumber \\
& \quad+\big(-5040 \eta ^3+190960 \eta ^2+137760 \pi ^2 \eta -7307920 \eta +107520\big) e^2\big) \cos ^2(u)-761600 \eta +\big(\big(-2240 \eta ^3-168000 \eta ^2 \nonumber \\
& \quad -424480 \eta \big) e^9+\big(28560 \eta ^3+242480 \eta ^2+34440 \pi ^2 \eta -1340224 \eta +725760\big) e^7+\big(-33040 \eta ^3-754880 \eta ^2 \nonumber \\
& \quad -172200 \pi ^2 \eta +5458480 \eta -221760\big) e^5+\big(40880 \eta ^3+738640 \eta ^2+30135 \pi ^2 \eta +1554048 \eta -2936640\big) e^3 \nonumber \\
& \quad +\big(-560 \eta ^3-100240 \eta ^2-43050 \pi ^2 \eta +3284816 \eta -389760\big) e\big) \cos (u)+\sqrt{1-e^2} \big(\big(\big(-127680 \eta ^2+544320 \eta  \nonumber \\
& \quad -739200\big) e^7+\big(-53760 \eta ^2-8610 \pi ^2 \eta +674240 \eta -67200\big) e^5\big) \cos ^5(u)+\big(\big(161280 \eta ^2-477120 \eta +537600\big) e^8 \nonumber \\
& \quad +\big(477120 \eta ^2+17220 \pi ^2 \eta -2894080 \eta +2217600\big) e^6+\big(268800 \eta ^2+25830 \pi ^2 \eta -2721600 \eta  \nonumber \\
& \quad +1276800\big) e^4\big) \cos ^4(u)+\big(\big(-524160 \eta ^2+1122240 \eta -940800\big) e^7+\big(-873600 \eta ^2-68880 \pi ^2 \eta +7705600 \eta  \nonumber \\
& \quad -3897600\big) e^5+\big(-416640 \eta ^2-17220 \pi ^2 \eta +3357760 \eta -3225600\big) e^3\big) \cos ^3(u)+\big(\big(604800 \eta ^2-504000 \eta \nonumber \\
& \quad  -403200\big) e^6+\big(1034880 \eta ^2+103320 \pi ^2 \eta -11195520 \eta +5779200\big) e^4+\big(174720 \eta ^2-17220 \pi ^2 \eta -486080 \eta  \nonumber \\
& \quad +2688000\big) e^2\big) \cos ^2(u)+\big(\big(-282240 \eta ^2-450240 \eta +1478400\big) e^5+\big(-719040 \eta ^2-68880 \pi ^2 \eta +8128960 \eta   \nonumber \\
& \quad  -5040000\big) e^3+\big(94080 \eta ^2+25830 \pi ^2 \eta -1585920 \eta -470400\big) e\big) \cos (u)-67200 \eta ^2+761600 \eta +e^4 \big(40320 \eta ^2  \nonumber \\
& \quad +309120 \eta -672000\big)+e^2 \big(208320 \eta ^2+17220 \pi ^2 \eta -2289280 \eta +1680000\big)-8610 \eta  \pi ^2 -201600\big)+8610 \eta  \pi ^2 \nonumber \\
& \quad+201600
 \bigg ] \, .
\end{align}
}
\clearpage
The 3~PN Kepler equation is 
{\small
\begin{align}
l &= l_\mathrm{0PN} + l_\mathrm{2PN} x^2 + l_\mathrm{3PN} x^3  + \mathcal{O}(x^4) \\
l_\mathrm{0PN} &=  u - e \sin u \\
l_\mathrm{2PN} &= \frac{1}{8\sqrt{1-e^2}(1-e \cos(u))} \left [ 
-12 (2 \eta -5) (u-v) (e \cos (u)-1)-e \sqrt{1-e^2} (\eta -15) \eta  \sin (u)
 \right ] \\
l_\mathrm{3PN} &= \frac{1}{6720 (1-e^2)^{3/2}(1-e \cos(u))^3} \bigg[ 
35 \big(96 \big(11 \eta ^2-29 \eta +30\big) e^2+960 \eta ^2+\eta  \big(-13184+123 \pi ^2\big) \nonumber \\
&\quad +8640\big) (u-v) (e \cos (u)-1)^3+3360 \big(-12 (2 \eta -5) (u-v)+12 e (2 \eta -5) \cos (u) (u-v) \nonumber \\
&\quad +e \sqrt{1-e^2} (\eta -15) \eta  \sin (u)\big) (e \cos (u)-1)^2+e \sqrt{1-e^2} \big(140 \big(13 e^4-11 e^2-2\big) \eta ^3-140 \big(73 e^4-325 e^2+444\big) \eta ^2 \nonumber \\
&\quad +\big(3220 e^4-148960 e^2-4305 \pi ^2+143868\big) \eta +e^2 \big(1820 \big(e^2-1\big) \eta ^3-140 \big(83 e^2+109\big) \eta ^2-\big(1120 e^2+4305 \pi ^2  \nonumber \\
&\quad +752\big) \eta +67200\big) \cos ^2(u)-2 e \big(1960 \big(e^2-1\big) \eta ^3+6720 \big(e^2-5\big) \eta ^2+\big(-71820 e^2-4305 \pi ^2+69948\big) \eta \nonumber \\
&\quad  +67200\big) \cos (u)+67200\big) \sin (u)
 \bigg]
\end{align}
}
where, as in Ref.~\cite{2006PhRvD..73l4012K}, we use 
{\small
\begin{align}
v - u = 
2 \tan ^{-1}\left(\frac{\sin (u) \beta _{\phi }}{1-\cos (u) \beta _{\phi }}\right)
\end{align}
}
and 
{\small
\begin{align}
\beta_\phi = 
\frac{1-\sqrt{1-e_{\phi }^2}}{e_{\phi }}
 \, .
\end{align}
}
$e_\phi$ is given by
{\small
\begin{align}
e_\phi &= e + e_{\phi\mathrm{1PN}} x + e_{\phi\mathrm{2PN}}x^2 + e_{\phi\mathrm{3PN}}x^3 + \mathcal{O}(x^{4}) \\
e_{\phi \mathrm{1PN}} &= 
-e (\eta -4)
 \\
e_{\phi \mathrm{2PN}} &= 
\frac{e}{96 \big(e^2-1\big)}
  \bigg [ 
\big(41 \eta ^2-659 \eta +1152\big) e^2+4 \eta ^2+68 \eta +\sqrt{1-e^2} (288 \eta -720)-1248
 \bigg ] \\
e_{\phi \mathrm{3PN}} &= 
-\frac{e}{26880 \big(1-e^2\big)^{5/2}}
  \bigg [ 
\big(13440 \eta ^2+483840 \eta -940800\big) e^4+\big(255360 \eta ^2+17220 \pi ^2 \eta -2880640 \eta +2688000\big) e^2 \nonumber \\
&\quad -268800 \eta ^2+2396800 \eta +\sqrt{1-e^2} \big(\big(1050 \eta ^3-134050 \eta ^2+786310 \eta -860160\big) e^4+\big(-18900 \eta ^3+553980 \eta ^2 \nonumber \\
&\quad +4305 \pi ^2 \eta -1246368 \eta +2042880\big) e^2+276640 \eta ^2+2674480 \eta -17220 \eta  \pi ^2-1451520\big)-17220 \eta  \pi ^2 \nonumber \\
&\quad -1747200
 \bigg ] \, .
\end{align}
}
This completes the expressions used in the conservative dynamics.  The
radiation reaction is given to 2~PN order in
Ref.~\cite{2006PhRvD..73l4012K} in terms of $n$ and $e_t$.  We again
substitute for $n$ in terms of $x$, and obtain
{\small
\begin{align}
M \dot x &= \dot x_\mathrm{0PN} x^5 + \dot x_\mathrm{1PN} x^{6}
 + \dot x_\mathrm{1.5PN} x^{13/2} + \dot x_\mathrm{2PN} x^{7} + \mathcal{O}(x^{15/2}) \\
\dot x_\mathrm{0PN} &= 
\frac{2 \big(37 e^4+292 e^2+96\big) \eta }{15 \big(1-e^2\big)^{7/2}}
 \\
\dot x_\mathrm{1PN} &= \frac{\eta}{420 (1-e^2)^{9/2}} \left [ 
-(8288 \eta -11717) e^6-14 (10122 \eta -12217) e^4-120 (1330 \eta -731) e^2-16 (924 \eta +743)
 \right ] \\
\dot x_\mathrm{1.5PN} &=  
\frac{256}{5} \eta  \pi  \kappa _E(e)
\end{align}
}
{\small
\begin{align}
\dot x_\mathrm{2PN} &=  \frac{\eta}{45360 (1-e^2)^{11/2}} \bigg [ 
\big(1964256 \eta ^2-3259980 \eta +3523113\big) e^8+\big(64828848 \eta ^2-123108426 \eta +83424402\big) e^6 \nonumber \\
&\quad +\big(16650606060 \eta ^2-207204264 \eta +783768\big) e^4+\big(61282032 \eta ^2+15464736 \eta -92846560\big) e^2+1903104 \eta ^2 \nonumber \\
&\quad +\sqrt{1-e^2} \big((2646000-1058400 \eta ) e^6+(64532160-25812864 \eta ) e^2-580608 \eta +1451520\big)+4514976 \eta \nonumber \\
&\quad -360224
 \bigg ] \, ,
\end{align}
}
for $\dot x$, and 
{\small
\begin{align}
M \dot e &= \dot e_\mathrm{0PN} x^4 + \dot e_\mathrm{1PN} x^{5} 
+ \dot e_\mathrm{1.5PN} x^{11/2} + \dot e_\mathrm{2PN} x^{6} + \mathcal{O}(x^{13/2}) \\
\dot e_\mathrm{0PN} &= 
-\frac{e \big(121 e^2+304\big) \eta }{15 \big(1-e^2\big)^{5/2}}
 \\
\dot e_\mathrm{1PN} &= 
\frac{e \eta }{2520 \big(1-e^2\big)^{7/2}}
  \bigg [ 
(93184 \eta -125361) e^4+12 (54271 \eta -59834) e^2+8 (28588 \eta +8451)
 \bigg ] \\
\dot e_\mathrm{1.5PN} &= 
\frac{128 \eta  \pi }{5 e}
  \bigg [ 
\big(e^2-1\big) \kappa _E(e)+\sqrt{1-e^2} \kappa _J(e)
 \bigg ] \\
\dot e_\mathrm{2PN} &= 
-\frac{e \eta }{30240 \big(1-e^2\big)^{9/2}}
  \bigg [ 
\big(2758560 \eta ^2-4344852 \eta +3786543\big) e^6+\big(42810096 \eta ^2-78112266 \eta +46579718\big) e^4 \nonumber \\
&\quad +\big(48711348 \eta ^2-35583228 \eta -36993396\big) e^2+4548096 \eta ^2+\sqrt{1-e^2} \big((2847600-1139040 \eta ) e^4+(35093520 \nonumber \\
&\quad -14037408 \eta ) e^2-5386752 \eta +13466880\big)+13509360 \eta -15198032
 \bigg ] \, ,
\end{align}
}
for $\dot e$.  These equations are written in terms of the functions
$\kappa_E$ and $\kappa_J$, given in Ref.~\cite{Damour:2004bz} in terms
of infinite sums of Bessel functions.  We reproduce them here for
completeness.
{\small
\begin{align}
\kappa_E &= 
\sum _{p=1}^{\infty } \frac{1}{4} p^3 \big(\big(\big(-e^2-\frac{3}{e^2}+\frac{1}{e^4}+3\big) p^2+\frac{1}{3}-\frac{1}{e^2}+\frac{1}{e^4}\big) J_p(p e){}^2+\big(-3 e-\frac{4}{e^3}+\frac{7}{e}\big) p J_p'(p e) J_p(p e) \nonumber \\
&\quad +\big(\big(e^2+\frac{1}{e^2}-2\big) p^2+\frac{1}{e^2}-1\big) J_p'(p e){}^2\big)
\\
\kappa_J &= 
\sum _{p=1}^{\infty } \frac{1}{2} p^2 \sqrt{1-e^2} \big(\big(-\frac{2}{e^4}-1+\frac{3}{e^2}\big) p J_p(p e){}^2+\big(2 \big(e+\frac{1}{e^3}-\frac{2}{e}\big) p^2-\frac{1}{e}+\frac{2}{e^3}\big) J_p'(p e) J_p(p e)+2 \big(1-\frac{1}{e^2}\big) p J_p'(p e){}^2\big)
\end{align}
}
These are functions of $e$ only, and are computed numerically using a
sufficient number of terms in the summation that the result converges
to within machine precision ($10^{-15}$).  For computational
efficiency, the resulting function is converted into an interpolating
polynomial, and the interpolation error is estimated to be $\sim
10^{-12}$ in the range $0 < e \le 0.4$.

\end{widetext}

\bibliography{references}

\end{document}